\documentclass[prb,aps,showpacs,byrevtex,amsmath,amssymb,preprintnumbers,twocolumn]{revtex4}
\usepackage{graphicx}

\begin{document}
%\graphicspath{{./plotsPY_Xa=0.10-out/}{./plots_gr/}}

%\preprint{APS/123-QED}
%\preprint{PREPRINT}
\newcommand{\Cmm}{{C/m$^2$}}
\newcommand{\etal}{{\em et al.\/}}
\newcommand{\romd}{{\rm d}}

\title{Population inversion of a NAHS mixture adsorbed into a cylindrical pore}
%\title{Entropy reservoirs in confined fluids: Population Inversion and Wetting Transitions}
%AUTHORS
%1st Author
\author{Felipe Jim\'enez-\'Angeles}
%\email{fangeles@imp.mx}
\affiliation{Programa de Ingenier\'{\i}a
Molecular, Instituto Mexicano del Petr\'oleo, L\'azaro C\'ardenas
152, 07730 M\'exico, D. F., M\'exico}
%
%2nd Author
\author{Yurko Duda}
%\email{ydouda@imp.mx}
\affiliation{Programa de Ingenier\'{\i}a
Molecular, Instituto Mexicano del Petr\'oleo, L\'azaro C\'ardenas
152, 07730 M\'exico, D. F., M\'exico}
%
%3rd Author
\author{Gerardo Odriozola}
%\email{godriozo@imp.mx}
\affiliation{Programa de Ingenier\'{\i}a
Molecular, Instituto Mexicano del Petr\'oleo, L\'azaro C\'ardenas
152, 07730 M\'exico, D. F., M\'exico}
%
%4th Author
\author{Marcelo Lozada-Cassou}
%\email{marcelo@imp.mx}
\affiliation{Programa de Ingenier\'{\i}a
Molecular, Instituto Mexicano del Petr\'oleo, L\'azaro C\'ardenas
152, 07730 M\'exico, D. F., M\'exico}
\date{\today{}}
\begin{abstract}
A cylindrical nanopore immersed in a non-additive hard sphere
binary fluid is studied by means of integral equation theories and
Monte Carlo simulations. It is found that at low and intermediate
values of the bulk total number density the more concentrated bulk
species is preferentially absorbed by the pore, as expected.
However, further increments of the bulk number density lead to an
abrupt population inversion in the confined fluid and an entropy
driven prewetting transition at the outside wall of the pore.
These phenomena are a function of the pore size, the
non-additivity parameter, the bulk number density, and particles
relative number fraction. We discuss our results in relation to
the phase separation in the bulk.
\end{abstract}
\pacs{61.30.Hn,61.46.Fg} \maketitle

\section{Introduction}

Fluids in confinement appear in many disciplines ranging from
biophysics to material science. Nanotubes, molecular channels
formed by transmembrane proteins, micelles, micropores and
nanopores in rocks are some examples of confining structures of
nanometric dimensions. Hence, understanding the effect of
confinement on fluids physicochemical properties is relevant for a
broad variety of technological areas such as catalysis, oil
recovery, and drugs delivery \cite{DNA}, to cite a few. Some
examples of phenomena associated with confinement are capillary
condensation \cite{Gubbins03}, new phases of water
\cite{waterconfined2} and wetting-drying transition \cite{Evans}.
In charged complex fluids, confinement may produce the formation
of a 2-D array stack of polyelectrolytes
\cite{Safinya1,OdriozolaPRL2006} and closed spherical and
cylindrical charged nanopores induce charge separation of a
confined electrolyte \cite{separation,Eloy}.

Capillary condensation and wetting are, perhaps, the most studied
phase transitions of confined fluids \cite{Tarazona87,EvansJCP86}.
When a vapor is confined, e.g. by a nanopore, in the appropriate
conditions, it condenses, such that this liquid is in equilibrium
with the bulk vapor: This phenomenon is known as capillary
condensation.  On the other hand, the fluid at a fluid-wall
interface may present two different regimes of adsorption: By
increasing the fluid number density, i.e., as the fluid approaches
to its saturation density, $\rho_{\rm sat}$, the adsorbed fluid
forms droplets (non-wetting) or spreads (wetting) on the surface,
producing a film of finite thickness of a condensed phase, i.e.,
1) In the non-wetting regime the adsorbed fluid forms a thin
layer, which increases slightly as the fluid approaches to
$\rho_{\rm sat}$. At $\rho_{\rm sat}$ the fluid condenses,
producing a macroscopic droplet, thus giving rise to a sudden
increase of adsorption. 2) At the prewetting regime, a thick layer
of fluid coexists with a thin one, competing to cover the wall as
the fluid density goes to $\rho_{\rm sat}$. At $\rho_{\rm sat}$ a
thick layer of fluid covers the wall, i.e., wetting.

%which end up at a critical point. Above  the wetting critical
%point, the fluid thickness increases in a continuous way as the
%pressure goes to $P_{\rm sat}$.

Capillary condensation and wetting have been widely studied in
pure simple fluids with fluid-fluid and confining wall-fluid
attractive interactions
\cite{Tarazona87,Evans86,EvansJCP86,GubinsMP87,ZhaoPRE07,GubbinsJCSF86,SwolJCSF86}.
On the other hand, entropy driven wetting and capillary
transitions of hard-core interacting fluids have been addressed
just recently \cite{DijkstraJPCM05}. Capillary phenomena for
confined complex fluids have been also addressed through the
Asakura-Oosawa (AO) model
\cite{DijkstraJPCM04a,BinderPRE06,LowenJPCM08}. In this model it
is considered a mixture of hard-sphere colloids and point
particles and has been used to study capillary condensation
\cite{DijkstraJPCM04a,BinderPRE06} and evaporation
\cite{RothJPCM06,SchmidtJPM04}.

Non additive hard sphere (NAHS) fluids have been studied in the
past for different confinement conditions
\cite{duda2003,duda2001,duda2004}. It is well known that this
model fluid exhibits a two phase separation in bulk \cite{Amar89},
hence, capillary and surface phase transitions, inside and outside
of a nanopore, are expected upon approaching to the bulk
coexistence curve. Here we will focus on the confinement induced
phase transitions of a NAHS binary mixture inside and outside a
cylindrical nanopore (athermal and exclusively driven by entropy).
We compare these two confinement induced phase transition diagrams
with that for a bulk fluid. It should be pointed out that the NAHS
model separates into two {\em dense} phases: one phase rich in
species A and poor in species B, while the other is oppositely
composed, i.e., rich in species B and poor in species A. On the
other hand, in confinement capillary condensation and evaporation
occur simultaneously. Thus, we will refer to the capillary induced
phase transition reported here as population inversion since the
phases are equally dense and oppositely composed.

\section{Model and Theory}

\begin{figure}[!ht]
{\includegraphics[width=7.5cm]{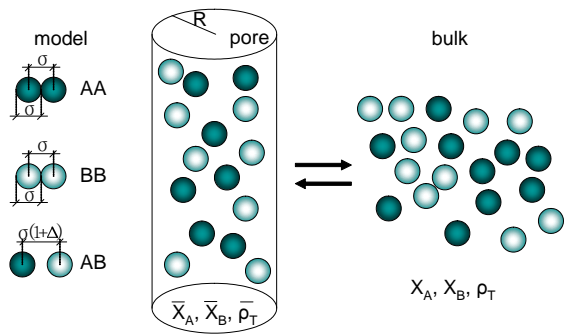}} \caption{Set up of the
NAHS model and the system under study. Dark and light gray spheres
represent species $A$ and $B$, respectively.} \label{scheme}
\end{figure}

The interest in NAHS is motivated by the observed nonadditivity in
real mixtures \cite{Bllemans} and by the belief that large
nonadditivity plays an essential role in the structure of some
liquid alloys \cite{gazzillo90}. The limiting case of the NAHS
model (so-called, Widom Rowlinson penetrable hard sphere mixture)
can be used to describe condensation phenomena
\cite{WidomJCP,MalijevskyJCP}. NAHS pair potential has been
successfully applied to the study of the morphology of composite
polymer particles \cite{DudaLng05}, the solubility of molecular
additives in different solvents \cite{DudaJPCB05}, the
microstructure of micelles \cite{Hamad04}, and for the description
of gas-gas phase transition at high pressure observed in mixtutes
of rare gases \cite{Schouten89}. Here we use the NAHS as a simple
way of modelling affinity between particles of the same species
and phobicity for particles of a different one
\cite{duda2001,duda2003,Yethiraj03,haro2005,lomba2005,duda2004,saija2007},
{such as, for example, a colloidal phase of oil interacting with a
water soluble polymer.} The model distinguishes between two
species (A and B) by defining a closest approach distance between
two particle of species $i$ and $j$, $\sigma_{ij}$, as follows:

\begin{eqnarray}
\sigma_{AA}=\sigma_{BB}= \sigma \nonumber \\
\sigma_{AB}= \sigma(1+\Delta)
\end{eqnarray}
being $\sigma$ the diameter of the particle, taken as the unit
length, and $\Delta$ the non additivity parameter. In bulk, for
$\Delta$$>$0 the mixture may separate in two phases, each of which
consists predominantly of one species, A or B. For the sake of
simplicity we consider here the symmetrical case of the NAHS
model, although the main conclusions of our work qualitatively do
not depend on the size asymmetry. The main features of the NAHS
model are depicted in Fig.~\ref{scheme}. Note that this model is
{\em athermal} since we are dealing with hard sphere interactions.

\begin{figure}[!ht]
{\includegraphics[width=6cm]{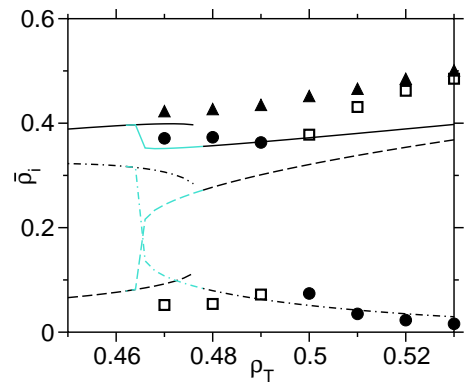}} \caption{Absorption curves
($\bar{\rho}_i$, with $i=$A,B,T) as a function of $\rho_T$, for
$R=3.5$, $\Delta=0.2$ and $X_A=0.06$: $\square$, $\bullet$ and
$\blacktriangle$ symbols are, respectively, MC results for
$\bar{\rho}_A$, $\bar{\rho}_B$ and $\bar{\rho}_T$, whereas the
dashed, dot-dashed and solid lines are the corresponding results
from HNC/PY theory. Black lines are obtained by increasing
$\rho_T$, whereas the light branches show a hysteresis interval
obtained upon decreasing $\rho_T$ from values above the upper
transition.} \label{inversion}
\end{figure}

\begin{figure}[!ht]
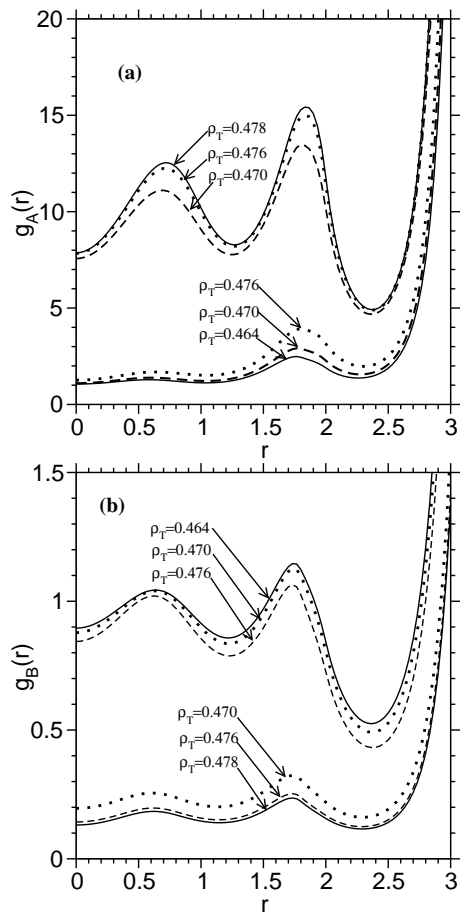

{\includegraphics[width=6cm]{FIG3a.eps}}
{\includegraphics[width=6cm]{FIG3b.eps}} \caption{Theoretical
reduced concentration profiles inside the cylindrical nanopore,
$g_A(r)$ and $g_B(r)$, for species A and B, respectively; and for
$R=3.5$, $\Delta=0.2$, $X_A=0.06$, and $\rho_T=$0.64, 0.70, 0.76,
0.78. The concentration profiles for $\rho_T=$0.70 and 0.76
correspond with the two branches of $\bar{\rho}_i$ vs $\rho_T$, in
Fig. \ref{inversion}, according to the color line.} \label{gsr}
\end{figure}

Let us consider the absorption of a binary mixture of NAHS
particles by a cylindrical pore of radius $R$ and infinite length,
immersed in a symmetrical NAHS bulk fluid. Hence, the confined
fluid and the infinite reservoir are at the same chemical
potential . The two species $A$ and $B$, are at a number density
$\rho_i$ ($i=A,B$), with the fluid total number density
$\rho_T=\rho_A+\rho_B$, and particles relative fraction
$X_i=\frac{\rho_i}{\rho_T}$. The relative fraction for species $i$
inside the pore is defined as
$\bar{X}_i=\frac{\bar{\rho}_i}{\bar{\rho}_T}$, being
$\bar{\rho}_T=\bar{\rho}_A+\bar{\rho}_B$, where $\bar{\rho}_i$ is
the average number density of species $i$ inside the pore (see
Fig.~\ref{scheme}). Note that because of the model symmetry, only
the interval $0\le X_A\le0.5$ needs to be considered. In this
work, this system is studied by two methods: Integral equation
theory and Monte Carlo (MC) simulations.

\begin{figure}[!ht]
\mbox{\put(0,165.9){\includegraphics[width=7cm]{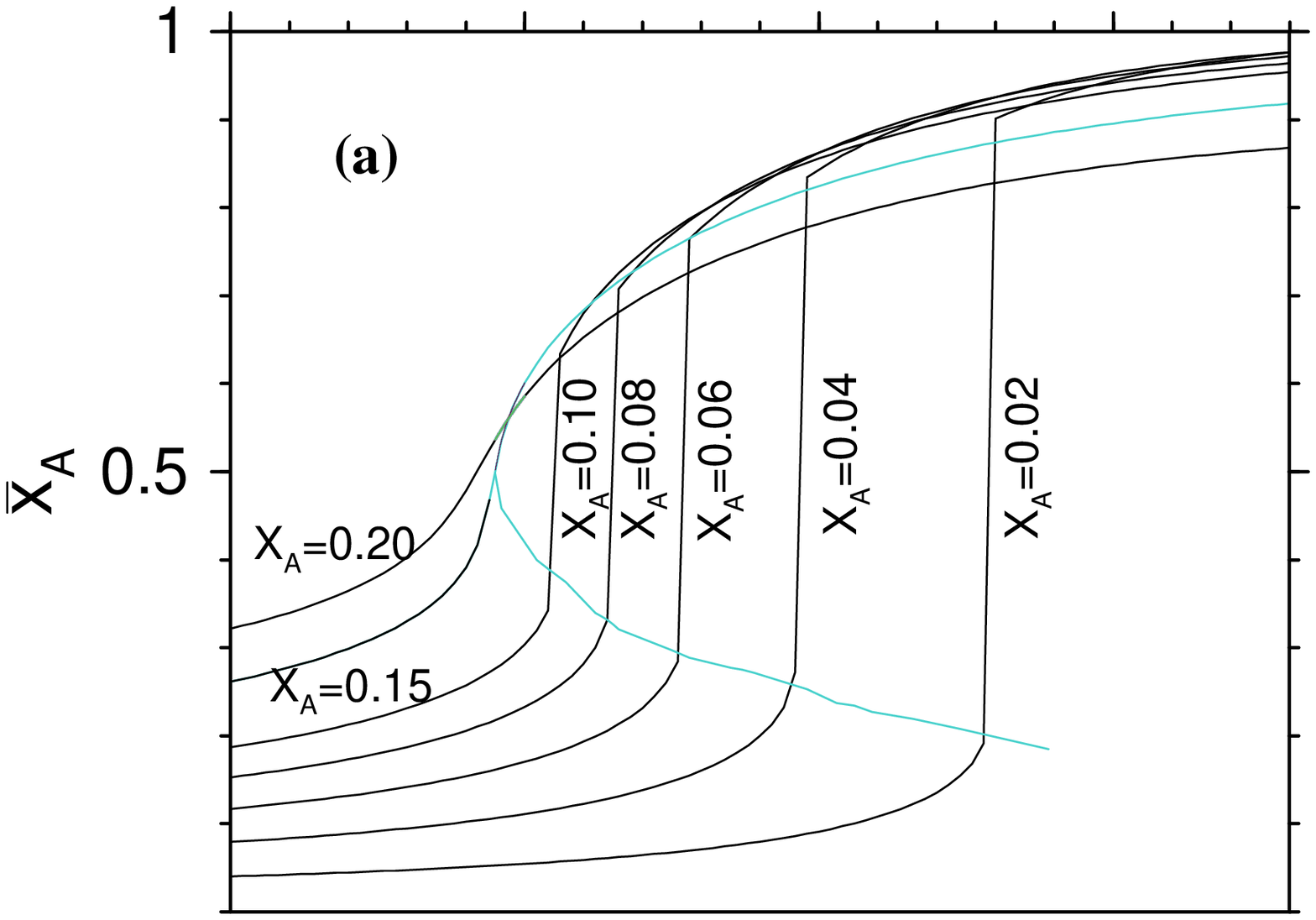}}
{\includegraphics[width=7cm]{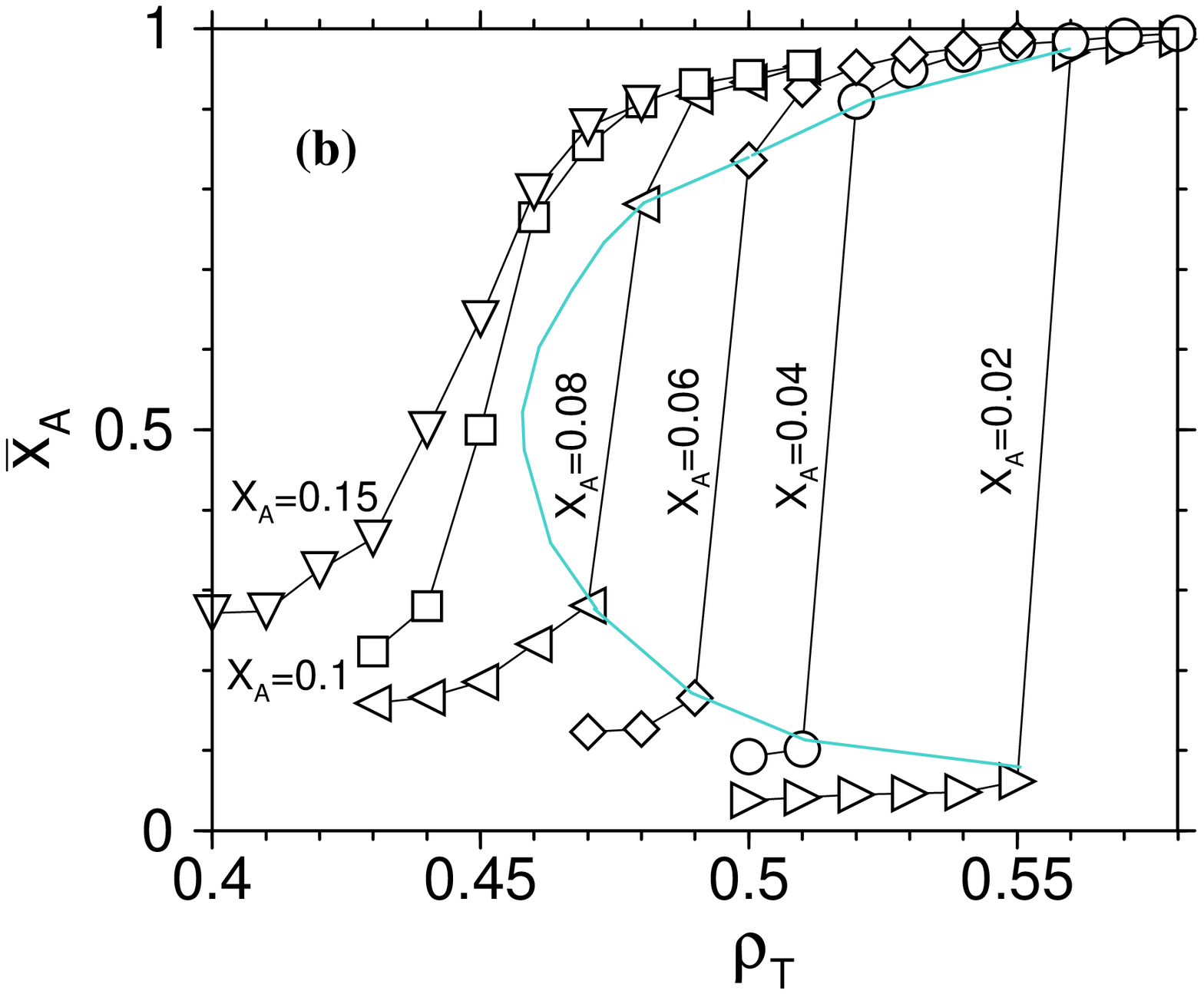}}} \caption{Relative A
fraction, $\bar{X}_A$, as a function of $\rho_T$, for $R=3.5$,
$\Delta=0.2$ and for the indicated bulk relative fraction, $X_A$:
(a)  from HNC/PY theory and (b) from MC simulations.}
\label{adsorp-Th}
\end{figure}

Liquid theories based on the Ornstein-Zernike integral equation
have been applied to both, charged and uncharged fluids confined
by planar, cylindrical and spherical nanopores, showing, in
general, good agreement with computer simulations
\cite{Alejandre,separation,duda2001,duda2004}. The method we use
to derive integral equations for confined fluids is based on the
equivalence between particles and fields \cite{henderson92a}.
Thus, the Ornstein-Zernike equation for inhomogeneous fluids is
obtained by considering the cylindrical pore as one more fluid
species ($\alpha$) at infinite dilution ($\rho_\alpha\to 0$). From
this inhomogeneous fluid integral equation, the hypernetted
chain/Percus-Yevick (HNC/PY) can be readily derived for a two
species NAHS fluid confined by a cylindrical pore, which relates
the reduced concentration profile of species $i$ inside the
cylinder, $g_{\alpha i}(r)$, the total correlation function,
$h_{\alpha i}(r)\equiv g_{\alpha i}(r)-1$; and the direct
correlation function, $c_{mi}(s)$, i.e.,

%for the nonadditive hard sphere mixture confined by the cylindrical pore
%
\begin{equation}
g_{\alpha i}(r)=\exp\left\{-\beta u_{\alpha i}(r)+
\sum_{m=1}^2\rho_m \int h_{\alpha m}(r') c_{mi}(s) d{\bf
r}'\right\}\label{eq:hnc}
\end{equation}
where $r$ and $r'$ are two radial cylindrical coordinates with its
origin at the center of the pore, $s=|{\bf r}-{\bf r}'|$ is the
relative distance between two particles of species $i$ and $m$ at
${\bf r}$ and ${\bf r}'$, respectively, and $d{\bf r}'$ is the
volume element; $\rho_m$ is the bulk number density of species
$m$; $\beta=1/(k_BT)$, where $k_B$ is the Boltzmann constant and
$T$ is the absolute temperature; $u_{\alpha i}(r)$ is the direct
interaction potential between a particle of species $i$ and the
pore, which is simply a hard-core interaction potential being
$u_{\alpha i}(r)=\infty$ for $r \ge R-\sigma/2$; $c_{m i}(s)$ is
approximated by the Percus-Yevick (PY) closure \cite{mcquarrie}.
The mean number density of species $i$ inside the pore,
$\bar{\rho}_i$, is computed as

\begin{equation}
\bar{\rho}_i = \frac{\bar{N}_i}{V_{p}} = \frac{2
\rho_i}{R^2}\int_0^{R-\sigma/2}g_{\alpha i}(r)rdr
\end{equation}
being $\bar{N}_i$$=$$2\pi L \rho_i\int_0^{R-\sigma/2}g_{\alpha
i}(r)rdr$ the number of particles of species $i$ absorbed into the
pore, $V_{p}=\pi L R^2$ the pore's volume, and $L$ the pore's
length. For the bulk fluid the HNC/PY equation is derived simply
by considering the $\alpha$ species particle to be equal to one of
the fluid species, say, $\alpha=$A or $\alpha=$B. For the bulk
fluid, it is interesting to note that there is a region of the
($X_A$, $\rho_T$) space where HNC/PY bifurcates into two solutions
and in a hysteretical cycle, such that the cycle region is within
the coexistence and spinodal curves.

MC simulations are performed in the modified Gibbs ensemble
\cite{panag1987}. We considered the NAHS mixture in the bulk cubic
simulation cell and absorbed inside the cylindrical pore, see
Fig.~\ref{scheme}. Thus the absorbed and bulk fluids are in
equilibrium. Unlike the usual Gibbs ensemble simulation, the
relative fraction ($X_A$) and bulk number density ($\rho_T$) of
the NAHS fluid are fixed, hence, we only perform creation and
annihilation of particles in the nanopore according with bulk
conditions. The length of the bulk box side and the cylinder axis
are large enough to avoid size effects. The number density and
relative fraction of the absorbed fluid have been calculated and
analyzed by the histogram method. Standard semi-grand canonical MC
simulations \cite{MC_pase-sep} are performed for obtaining the
two-phase separation diagram of the bulk fluid. Our MC results for
the bulk are in agreement with those reported by Amar
\cite{Amar89}.

%\begin{figure}
%{\includegraphics[width=8cm]{adsorp-MC.eps}} \caption{Relative $A$
%concentration, $\bar{X}_A$, inside the pore as a function of
%$\rho_T$, for $R=3.5$, $\Delta=0.2$ and for different bulk
%relative concentrations, $X_A$. Data are obtained by means of MC
%simulations, where symbols $\square$, $\circ$, $\bigtriangleup$,
%$\bigtriangledown$, $\diamond$, $\triangleleft$, and
%$\triangleright$ correspond to $X_A$$=$0.40, 0.20, 0.15, 0.10,
%0.06, 0.04, and 0.02, respectively. The solid gray line is defined
%by the points of $\bar{X}_A(\rho_T,X_A)$ close to the
%discontinuities.} \label{adsorp-MC}
%\end{figure}

\section{Results}

\subsection{Inside the cylindrical pore: Population Inversion}

A cylindrical pore with $R$$=$3.5 and a NAHS fluid with
$\Delta=0.2$ are considered. {It should be pointed out, however,
that all the phenomena discussed here are observed whenever
$\Delta > 0$.} Fig.~\ref{inversion} shows the absorbed fluid
number densities $\bar{\rho}_A$, $\bar{\rho}_B$, and
$\bar{\rho}_T$, as a function of the bulk total density, $\rho_T$,
for a given relative A-particle fraction, $X_A$. For low values of
$\rho_T$ species B is preferably absorbed by the pore
($\bar{\rho}_B$$\gg$$\bar{\rho}_A$), as expected since $X_B=0.94$.
Interestingly, $\rho_A$ is lower than $\bar{\rho}_A$ for all
$\rho_T$, which is atypical for purely repulsive interactions
\cite{Alejandre}. For a certain transition bulk number density,
$\rho_T^t$, $\bar{\rho}_B$ has a discontinuous drop, while
$\bar{\rho}_A$ has the opposite behavior. Related phenomena have
been observed with the confined AO model
\cite{DijkstraJPCM04a,BinderPRE06,RothJPCM06,SchmidtJPM04}. In
that case, the transition can be explained by an effective
colloid-colloid attractive interaction originated by the presence
of polymers. Our case can be seen as a simultaneous A-species
condensation and B-species evaporation. Furthermore, no
preferential attraction is present because of the model symmetry.
We refer to this phenomenon as {\em population inversion} (PI),
i.e., for $\rho_T$$<$$\rho^t_T$, $\bar{\rho}_B$$\gg$$\bar{\rho}_A$
and for $\rho_T$$>$$\rho^t_T$, $\bar{\rho}_A$$\gg$$\bar{\rho}_B$,
{and, we believe, this phenomenon is related to i) the
catastrophic inversion of adsorbed water-oil emulsions
\cite{Sajjadi04} and ii) the selective adsorption of two nearly
similar liquids by a membrane \cite{StevenNat08}.}

%In bulk, this system exhibits a separation into a poor-in-colloids
%and rich-in-colloids phases, (similar to the gas-liquid
%transition). In confinement, the AO model fluid exhibits a
%capillary phase transition (characterized by an abrupt increment
%of the colloids concentration inside an slit-like pore) coexisting
%with a poor-in-colloids phase in bulk. These transitions can be
%explained by an effective colloid-colloid attractive interaction
%originated by the presence of polymers. Hence, this transition is
%referred to as capillary condensation of the AO model fluid, in
%analogy with the confinement induced phase transition for a simple
%liquid with attractive interactions.

The HNC/PY and MC results agree in this prediction. However the MC
results for $\bar{\rho}_T$ are
systematically higher than those from HNC/PY, %implying that the
%bulk MC chemical potential is higher than that obtained from HNC/PY,
due to the unavoidable approximations in many-body theories. More
interesting is that $\rho^t_T$ from HNC/PY  is lower than that
from MC. This is because HNC/PY absorption of A-particles is
higher than that of MC and, hence, there are more A-B pairs inside
the pore, {which in turn gives rise to a higher excluded volume
and this prompts the transition}: In the NAHS model  AB pairs have
higher excluded volume than AA or BB pairs. Notice that the MC
transition occurs when its $\bar{\rho}_A$ reaches around the same
value as that of HNC/PY at the transition.

Black lines in Fig.~\ref{inversion} are successive solutions of
Eq.~\eqref{eq:hnc} obtained by increasing $\rho_T$. On the other
hand, the light branches correspond to the solutions calculated by
decreasing $\rho_T$ from above the $\rho_T^t$ value. Notice, that
both solutions coincide in almost all the studied interval of
$\rho_T$, however there is an interval within which Eq.
\eqref{eq:hnc} has two solutions, i.e., an hysteresis cycle is
found. These two solutions imply the presence of metastable states
and two coexisting phases inside the cylindrical nanopore. MC
results qualitatively confirm the presence of hysteresis, although
the $\rho_T$ interval is narrower ($\Delta \rho_T$$\approx$0.01).

Fig.~\ref{gsr} shows the reduced concentration profiles obtained
from HNC/PY, $g_{\alpha i }(r)$, for both species ($i=$A, B),
inside the cylindrical nanopore. The solid lines are concentration
profiles around the PI transition. In Fig.~\ref{gsr}a, near
$\rho_T=0.464$, as we increase $\rho_T$ we see a sudden increase
in the species A population, inside the pore, when going from
$\rho_T=0.476$ to $\rho_T=0.478$. Because of the hysteretical
nature of the PI transition, when decreasing the bulk
concentration there is a sudden decrease of A species
concentration below $\rho_T=0.470$. In Fig~\ref{gsr} the
corresponding hysteretical behavior of the B species is seen. Of
course, an increase in species A implies a decrease in species B,
and viceversa. This concentration interval correspond to
hysteresis cycle shown in Fig.\ref{inversion}: The light branches
correspond to the decreasing concentration profiles in
Fig.~\ref{gsr}.
%Black curves are a set of solutions of Eq.~\eqref{eq:hnc} obtained
%by increasing the fluid total number density from $\rho_T=$0.64,
%whereas light curves are solutions for $\rho_T=$ 0.70 and 0.76
%obtained by decreasing $\rho_T$ from above 0.76.
Thus, for $\rho_T\lesssim$0.464 and $\rho_T\gtrsim$0.476 a single
solution is found, whereas, within this interval (0.464 $\lesssim
\rho_T \lesssim$ 0.476) two families of solutions are found.
Notice that only slight changes of $g_A(r)$ and $g_B(r)$ occur
before and after the jump, either by increasing or decreasing
$\rho_T$. The transition is seen by an increase (decrease) of
$g_A(r)$ ($g_B(r)$) of about an order of magnitude. Both species,
on the other hand, show a three-layer structure inside the
nanopore.

Fig.~\ref{adsorp-Th} shows results from HNC/PY integral equations
theory and MC simulations  for the relative fraction of species A
inside the cylinder, $\bar{X}_A$, as a function of the bulk total
number density, $\rho_T$, for a set of values of the A species
bulk relative fraction, $X_A$, such that $X_A$$<$$X_B$.  The light
curve defines the transition points. Notice that this curve {is
not a coexistence curve}, but defines an instability region. The
PI is observed only for $X_A$ below a critical value, $X^c_A$. The
critical isopleth predicted by MC simulations is
$X^c_A$$\approx$0.14 with a consolute point at
$\rho^{c}_T$$\approx$0.451, whereas integral equations predict
$X^c_A$$\approx$0.13 and $\rho^{c}_T$$\approx$0.46. Our analysis
of model parameters indicates that $X^c_A$ depends on $R$ and
$\Delta$ in a nontrivial way.

\begin{figure}[!ht]
{\includegraphics[width=8.0cm]{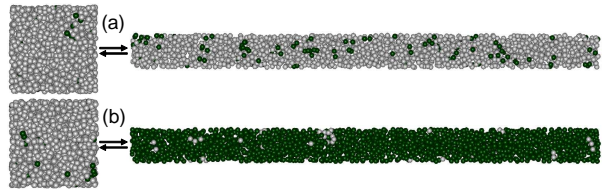}} \caption{Snapshots of
typical MC configurations for the confined fluid. (a)
$\rho_T=0.55\lesssim\rho^t_T$ and (b)
$\rho_T=0.56\gtrsim\rho^t_T$. Dark and light gray spheres
represent species $A$ and $B$, respectively. The left square boxes
are the bulk snapshots which are in equilibrium with their
respective confined fluids, in the cylindrical boxes. In both
cases $X_A=0.02$, $R=3.5$, and $\Delta=0.2$.} \label{snapshot}
\end{figure}

Fig.~\ref{snapshot} shows two snapshots, before and after the
population inversion transition, for $X_A=0.02$. Note that,
although the bulk number densities are very close in both cases,
the fluid inside the cylindrical pore dramatically changes its
composition: In (a) $\bar{\rho}_A=0.03$ and $\bar{\rho}_B=0.461$,
whereas in  (b) $\bar{\rho}_A=0.496$ and $\bar{\rho}_B=0.015$.

\begin{figure}[!ht]
{\includegraphics[width=6cm]{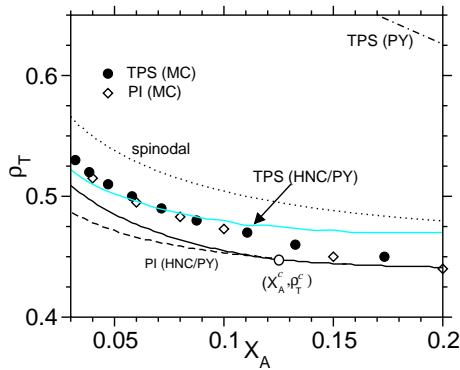}} \caption{Transition number
density for the bulk two-phase separation (TPS) and for the
population inversion (PI) of a NAHS fluid ($\Delta=0.2$) inside
the cylindrical nanopore($R=3.5$). The black circles and light
curve are the MC and HNC/PY results for the bulk TPS,
respectively. White diamonds and the black solid line are the MC
and HNC/PY PI diagrams, respectively. The black solid and dashed
lines defines the upper and bottom number density transitions,
i.e., the limits of hysteresis. The theoretical consolute point,
$X^c_A\approx$0.13, is defined by the point where the black solid
and dashed line converge. The dotted line is the HNC/PY bulk
spinodal, whereas the black dotted-dashed line is the bulk TPS
from PY integral equations theory. The PI consolute point
$(X_A^c,\rho_T^c)$ is signaled by an open circle.} \label{diagram}
\end{figure}
%In contrast the PY prediction for the demixing bulk diagram is quantitative different from the MC
%result (dotted-dashed line in Fig.~\ref{diagram}). { It should be pointed out that theory indicates
%that PI inside the cylinder corresponds to the bulk one-phase region suggesting that phase
%separation and PI are not directly related.}

The bulk NAHS fluid exhibits a fluid-fluid two phase separation
(TPS) \cite{Amar89}. In Fig.~\ref{diagram} we show MC (black
circles) and HNC/PY (solid light line) transition number
densities, $\rho_T^t$, for the bulk fluid, as a function of $X_A$.
Above the bulk TPS coexistence curve the fluid may separate into a
rich A-species phase and a rich B-species phase. Below the curve
the fluid is homogeneous. Between the coexistence (light) TPS and
spinodal curves (dotted) a metastable homogeneous fluid is found.
Also in Fig.~\ref{diagram} we show the MC and HNC/PY transition
number densities, $\rho_T^t$ as a function of $X_A$, at which
there is a PI inside the cylindrical pore, i.e.,
$\bar{X}_A>$$\bar{X}_B$, by increasing $\rho_T$, or
$\bar{X}_A<$$\bar{X}_B$, by decreasing $\rho_T$. From the HNC/PY
approach, the black solid and dashed lines are constructed with
the transition points at which PI occurs by increasing/decreasing
$\rho_T$,
%, whereas the dashed line correspond to the transition
%$\bar{X}_A<$$\bar{X}_B$ when $\rho_T$ is decreases
i.e., the hysteresis cycle pointed out in Figs.~\ref{inversion}
and \ref{gsr} is defined by the region between the solid and
dashed lines. These two lines converge at the consolute point,
i.e., $X^c_A$$\approx$0.13. For $X_A> X^c_A$, there is still a PI
but it happens in a continuous way as $\rho_T$ increases, i.e.,
there is not a sudden transition. According to the HNC/PY
approach, the PI curve is shifted towards lower $\rho_T$ values
respect to the TPS curve, i.e., as $\rho_T$ increases first occurs
a PI inside the pore and later a TPS in the bulk. A closer
inspection of MC data clearly confirm this prediction. The black
dot-dashed at the right upper corner is the PY transition curve
for the bulk TPS, and is included for reference.
%This trend is verified by MC simulations, although the differences
%in $\rho_T$ for EDWT and PI are much smaller.

\begin{figure}[!ht]
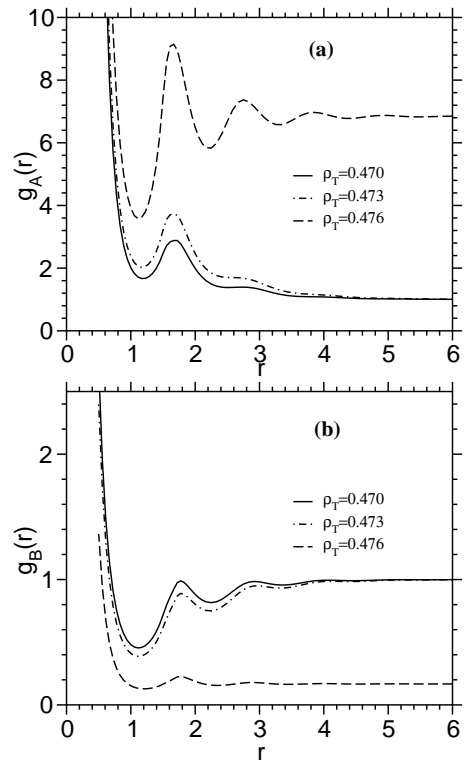

{\includegraphics[width=6cm]{FIG7a.eps}}
{\includegraphics[width=6cm]{FIG7b.eps}} \caption{Theoretical
reduced concentration profiles outside the cylindrical nanopore,
$g_A(r)$ and $g_B(r)$, for species A and B, respectively; and for
$R\to \infty$, $\Delta=0.2$, $X_A=0.10$, and $\rho_T=$0.470,
0.473, 0.476.} \label{gout}
\end{figure}

\begin{figure}[!ht]
{\includegraphics[width=6cm]{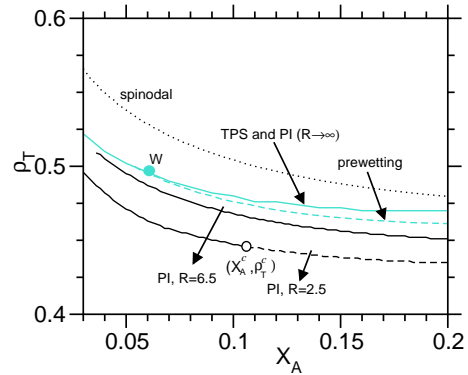}} \caption{Theoretical EDPW
transition curve for $R=\infty$ (dashed light) and PI curves for
$R=$2.5, 6.5 (solid black), and $\infty$ (solid light). Also is
plotted the bulk spinodal curve (black dotted), while the TPS
coexistence curve coincides with the PI for $R\to$$\infty$ (solid
light). The PI consolute point $(X_A^c,\rho_T^c)$ is signaled by
an open circle. The filled light circle is the converging point of
the prewetting and TPS curves. In all cases $\Delta=0.2$.}
\label{diagramHNC}
\end{figure}

Although not shown, for a fixed value of $R$, PI curves shift
towards lower bulk number densities as $\Delta$ increases, while
the consolute point is at lower $X_A$ values, responding to the
higher volume demand of the A-B coordinates. At constant $\Delta$,
the PI transition diagram is displaced to higher $\rho_T$ values
as $R$ increases, whereas the consolute point shifts towards
higher $X_A$ values, because there is more accessible volume
inside the pore.

\subsection{Outside the cylindrical pore: Wetting-like transition}

The NAHS fluid in contact with the outside surface of the
cylindrical nanopore was also studied by means of integral
equations. Fig. \ref{gout} shows the reduced concentration
profiles for both species, at the outside surface of the
cylindrical nanopore, for $\rho_T$=0.470 and 0.473 (below the bulk
TPS coexistence curve) and for $\rho_T$=0.476 (slightly above the
bulk TPS coexistence curve). For this inhomogeneous fluid, below
the TPS, we found a preferable adsorption of the less concentrated
A-particles, which like to be next to the cylinder. This can be
seen by the high and increasing value of $g_A(r)$ on the
cylinder's surface, and the sudden long range correlation of
particles A with respect to the cylinder surface, whereas $
g_B(r)$ decreases, as $\rho_T$ increases.
%since in this way the number of heterocoordinates is reduced and,
%thus, space is optimized favoring entropy maximization.
In order to evaluate this phenomenon we define the adsorption at
the cylinder outside surface as

\begin{equation}
\Gamma_i=2\pi\int_{R+a/2}^{\infty}\rho_i[g_i(r´)-1]r´dr´
\end{equation}
with $i=A,B$, the species label.

Above a certain value of $X_A=$ $X^{\rm w}_A$, we find a bulk
total number density, $\rho_T^{\rm pw}$ (below the PI and TPS
transition curves), for which the adsorption on the outside
cylinder's surface of A-particles is higher than for B-particles,
i.e., $\Gamma_A>\Gamma_B$. In Fig.~\ref{diagramHNC}, we
constructed the prewetting line with this criterion. This
phenomenon is similar to the prewetting occurring nearby the
liquid-gas coexistence which, by increasing $\rho_T$, becomes
wetting. Since there is not a surface-fluid attractive energetic
contribution, we refer to this phenomenon as an entropy driven
prewetting (EDPW). The dashed light curve in Fig. \ref{diagramHNC}
is the prewetting line for $R\to\infty$, which coincides with the
TPS bulk coexistence curve (solid light line) for $X_A<X^{\rm
w}_A\approx 0.065$. The converging point separates the regimes of
wetting ($X_A\ge$$X_A^W$) and nonwetting ($X_A<X_A^W$).

The limiting PI transition curve for $R$$\to$$\infty$ coincides
with the bulk TPS coexistence curve, which is reached about
$R=$$20$. As $R$ decreases the PI transition occurs at lower
$\rho_T$ values. {\em Hence, for a given cylinder size immersed in
the bulk, we first see the PI transition, then the EDPW
transition, and finally the TPS  transition, as $\rho_T$
increases.} Whether a PI or EDPW transition occur, the AB-pairs
adsorption inside the pore or at the outside cylinder's surface
implies that the number of AB-pairs decreases in the bulk, thereby
increasing the total accessible volume. For a constant $\rho_T$, a
larger accessible volume implies more entropy. Therefore, the
cylinder's cavity and its outside surface act as an entropy
reservoir.

As $R$ decreases the prewetting line shifts towards the TPS
coexistence curve, meanwhile its corresponding $X^{\rm w}_A$ is
displaced towards higher values of $X_A$ (not shown). For small
values of $R$, the prewetting lines are in general above their
corresponding PI transition curves (see Fig.~\ref{diagramHNC}),
except for large values of $R$ ($R>20$), where it is below its PI
transition. This behavior, which at first might sound
contradictory, can be understood if one realizes that the smaller
the cylinder the larger the confinement inside the pore, but the
smaller the confinement exerted by its outside surface, and that
confinement promotes the PI or EDPW transitions.

From the perspective of the pore cavity, given that the chemical
potential is the same outside and inside the cylinder, to have the
same energy necessary to bring a particle A from infinity to
inside or outside the pore, the inside average concentration must
be smaller than that outside, to compensate for their confinement.
This effect is clearly seen in Fig.~\ref{inversion}, in the
$\bar{\rho}_T$ curve as a function of $\rho_T$, and is magnified
at the PI transition, i.e., the number of A-B pairs increases
inside the pore after the PI transition, thereby exhibiting the
transition mechanism. Since the confinement is lower at the
outside pore surface the EDPW transition occurs at higher
$\rho_T$.

Trough PI and EDPW transitions,  the system adopts configurations
in which the number of AB pairs are decreased as a mechanism to
maximize accessible volume, i.e., entropy. That is, on one hand,
the asymmetric pairs (AB) occupy a higher volume than symmetric
pairs (AA or BB) and, on the other hand, particles next to a
surface (inside or outside the cylinder) have a lower coordination
number, both mechanisms implying an increase of accessible volume.
Hence, the system uses the cylindrical cavity and its outside
surface as an entropy reservoir.
%from where the fluid gains accessible volume.

%In general, the EDW is not as abrupt as the PI transition meaning
%that $\Gamma_A$$\approx$$\Gamma_B$ above the prewetting line
%whereas, in general, $\bar{\rho_A}$ and $\bar{\rho_B}$ are totally
%different above the PI transition, but they are close
%($\bar{\rho_A}$$\approx$$\bar{\rho_B}$) only close to the
%consolute point.
%PI and WT transitions are differentiated in that PI is prompted by
%confinement whereas WT is prompted by
\section{Conclusions}

We have studied the adsorption of a non-additive hard sphere
mixture on the inner and outer surfaces of a cylindrical nanopore.
Upon approaching towards the bulk two-phase separation (but well
inside the homogeneous phase) it was found a phase transition of
the confined fluid, referred to as population inversion. Such a
transition is identified by a sudden change in composition of the
confined fluid, involving an absorption (desorption) of the less
(more) concentrated species in bulk. Further increments of the
bulk total number density give raise to a transition at the
cylinder's outside surface, which is characterized by a preferable
adsorption of the less concentrated species in bulk over the more
concentrated one, referred to as entropy driven prewetting. All
these phenomena, confinement induced population inversion and
wetting-like transition are akin to the capillary
condensation-evaporation and wetting for simple fluids. However
two differences should be pointed out: 1) These novel effects
occur near the fluid-fluid two phase separation, whereas for
simple fluids the related phenomena occur near the gas-liquid
phase transition, and 2) These effects are ruled by entropy only.
The smaller cylinder the higher the confinement inside the
cylinder, but lower the confinement due to the outside surface of
the cylinder. The higher the confinement the lower the
concentration at which the PI and EDPW transitions occur. Hence, a
particular selection of particles and nano-cylinders sizes can be
used for selective adsorption, in different applications.

%\bibliographystyle{prsty}
%\bibliography{confinedsystems}

\end{document}